\documentclass[amssymb,twocolumn,prl,aps,showpacs,floatfix,letterpaper]{revtex4}
\usepackage{graphicx}
\usepackage{epsfig,color}
\begin{document}

\title{Emergence of a superconducting state from an antiferromagnetic phase in single crystals of the heavy fermion compound Ce$_{2}$PdIn$_{8}$}

\author{D. Kaczorowski, A. P. Pikul, D. Gnida, and V. H. Tran}
\affiliation{Institute of Low Temperature and Structure Research, Polish Academy of Sciences, P. O. Box 1410, 50-950 Wroc\l aw, Poland}

\begin{abstract}
Single crystals of Ce$_2$PdIn$_8$ were studied by means of magnetic susceptibility, electrical resistivity and specific heat measurements. The compound was found to be a heavy fermion clean-limit superconductor with $T_{\text{c}}$ = 0.68 K. Most remarkably, the superconductivity in this system emerges out of the antiferromagnetic state that sets in at $T_{\text{N}}$ = 10 K, and both cooperative phenomena coexist in a bulk at ambient pressure conditions.
\end{abstract}
\pacs{74.20.Mn;74.70.Tx;74.25.Fy;74.25.Bt;75.30.Mb}
\maketitle

\newpage
One of the most intriguing problems in the contemporary solid state physics concerns the mutual relation between the two major cooperative phenomena, i.e. magnetism and superconductivity (SC). Regarded for many years as definitive antagonists they are considered nowadays as close partners in strongly correlated electron systems like heavy fermion (HF) compounds, cuprates and ruthenates. The emerging SC is usually unconventional, both in terms of non-phononic coupling mechanism and symmetry of the pairing. Spectacular discoveries of pressure-induced SC in antiferromagnetically ordered CeRhIn$_5$ \cite{heg_cerhin5} and ambient-pressure HF superconductivity in paramagnetic CeCoIn$_5$ \cite{pet_cecoin5} and CeIrIn$_5$ \cite{pet_ceirin5} have ignited intense investigations of this family of compounds. These studies resulted in many crucial findings in regard to coexistence of antiferromagnetism (AF) and SC, magnetically-mediated SC, non-Fermi liquid (NFL) features near magnetic instability, formation of a Fulde-Farrell-Larkin-Ovchinnikov state, etc. In parallel, a series of structurally closely related indides Ce$_2$\emph{M}In$_8$ (\emph{M} = Co, Rh, Ir) was being studied and so far the SC state has been established for Ce$_2$CoIn$_8$ \cite{che_ce2coin8} and Ce$_2$RhIn$_8$ \cite{nic_ce2rhin8}. The physical properties of these phases were found very similar to those of the Ce\emph{M}In$_5$ phases, yet their characteristic energy scales are reduced, likely because of higher effective dimensionality of the Ce$_2$\emph{M}In$_8$ systems \cite{2d3d}.

Recently we reported on the intriguing properties of a novel representative of the Ce$_2$\emph{M}In$_8$ series, i.e. Ce$_2$PdIn$_8$ \cite{sces08}. This compound, studied in polycrystalline form, has been characterized as a paramagnetic HF system with a NFL character of the electronic ground state. Here, we report the discovery of HF SC in single crystals of Ce$_2$PdIn$_8$, which emerges out of an AF ordered state.

Single crystals of Ce$_2$PdIn$_8$ were grown from an In-flux, as described for Ce$_2$CoIn$_8$ \cite{che_ce2coin8}. X-ray diffraction and microprobe analysis proved the expected crystal structure (Ho$_2$CoGa$_8$-type), proper stoichiometry and good homogeneity of the individual specimens. Physical property measurements were carried out by standard experimental techniques using commercial equipment.

\begin{figure}[h]
\includegraphics[width=0.82\columnwidth]{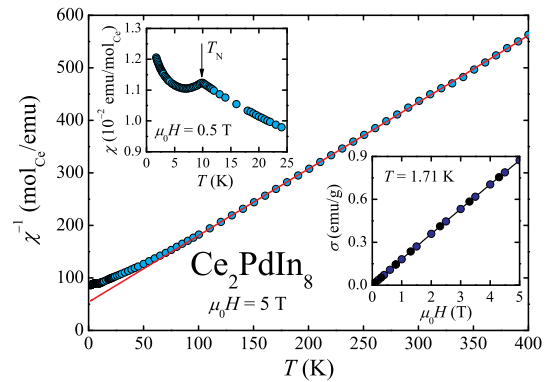}
\caption{\label{fig:Fig1} (color online) Temperature dependence of the reciprocal molar magnetic susceptibility of Ce$_2$PdIn$_8$ taken in magnetic field of 5 T. The solid line is the Curie-Weiss fit. Upper inset: low-temperature behavior of the magnetic susceptibility measured in $\mu_0 H$ = 0.5 T. Lower inset: field variation of the magnetization taken at 1.71 K with increasing (full circles) and decreasing (open circles) magnetic field strength.}
\end{figure}

Fig. \ref{fig:Fig1} displays the inverse magnetic susceptibility of Ce$_2$PdIn$_8$, measured on a specimen consisting of several small single crystals freely placed in a sample holder. Above about 80 K, $\chi(T)$ exhibits a Curie-Weiss behavior with the effective magnetic moment $\mu_{\text{eff}}$ = 2.51 $\mu_{\text{B}}$ and the Weiss temperature $\theta_{\text{p}}$ = -42.6 K. The experimental value of $\mu_{\text{eff}}$ is very close to that expected for a Ce$^{3+}$ ion. The large negative $\theta_{\text{p}}$ suggests the presence of Kondo interactions with the characteristic temperature $T_{\text{K}} \approx |\theta_{\text{p}}/4|$ \cite{hewson} of about 10 K. Deviation of the $\chi^{-1}$ vs. $T$ dependence from a straight-line behavior, observed below 80 K, may be attributed to crystalline electric field (CEF) effect. A maximum in $\chi(T)$ located at $T_{\text{N}} \approx$ 10 K (see the upper inset to Fig. \ref{fig:Fig1}) manifests a magnetic phase transition into an AF state. Moreover, in the ordered region, the susceptibility shows a distinct upturn that may hint at another phase transition occurring at $T <$ 1.71 K. The magnetization, measured at 1.71 K, is a linear function of the magnetic field strength with no hysteresis effect (see the lower inset to Fig. \ref{fig:Fig1}), in line with the AF character of the compound at this temperature.

\begin{figure}[h]
\includegraphics[width=0.77\columnwidth]{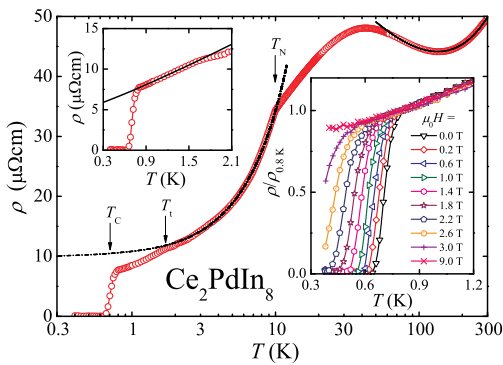}
\caption{\label{fig:Fig2} (color online) Temperature dependence of the electrical resistivity of single-crystalline Ce$_2$PdIn$_8$, measured perpendicular to the $c$-axis. The solid and dash-dotted line represent the fits discussed in the text. Upper inset: low-temperature resistivity with a linear scale. The solid line is the fit $\rho \sim T$. Lower inset: reduced resistivity in the vicinity of the superconducting transition, measured in zero and a few magnetic fields applied along the $c$-axis.}
\end{figure}

The temperature variation of the electrical resistivity of Ce$_2$PdIn$_8$, measured with the electrical current flowing within the tetragonal $a$-$b$ plane, is shown in Fig. \ref{fig:Fig2}. In the paramagnetic region, it reveals typical Kondo lattice behavior, with rather weak temperature dependence down to $T_{\text{max}}$ = 40 K, where $\rho(T)$ exhibits a broad maximum, and a rapid drop at lower temperatures. This maximum at $T_{\text{max}}$ likely arises due to a crossover from incoherent to coherent Kondo regime. Above about 65 K the experimental data can be well described by a function $\rho(T)=(\rho_0+\rho_0^\infty)+c_{\text{ph}}T+c_{\text{K}}\ln T$, with the parameters: $\rho_0+\rho_0^\infty$ = 97.3 $\mu \Omega$cm, $c_{\text{ph}}$ = 0.09 $\mu \Omega$cm/K and $c_{\text{K}}$ = -13.6 $\mu \Omega$cm. The first term is a sum of residual scattering on static defects in the lattice and scattering on disordered magnetic moments. The second term accounts for electron-phonon scattering processes, whereas the third one represents Kondo-type spin-flip scattering.

The most important feature of single-crystalline Ce$_2$PdIn$_8$ is a rapid drop of the resistivity to zero value that clearly manifests the onset of SC state. The critical temperature, defined as the temperature at which the resistivity drops to one tenth its value above the SC transition, is $T_{\text{c}}$ = 0.68 K. As displayed in the lower inset to Fig. \ref{fig:Fig2}, in applied magnetic fields SC gets suppressed, and $T_{\text{c}}$ gradually decreases with raising the field strength.

The magnetic phase transition at $T_{\text{N}}$ = 10 K manifests itself as a distinct kink in $\rho(T)$. At $T_{\text{t}}$ = 1.7 K, a less pronounced change in the slope of $\rho(T)$ is seen that might reflect the other phase transition anticipated from the upturn in $\chi(T)$. As $T_{\text{t}}$ depends on the magnetic field strength (it drops down to 1.37 K in $\mu_0 H$ = 9 T) but the magnitude of this feature hardly varies with field, one may suppose that it has magnetic origin (eg. it could manifest some change in the AF structure, as it occurs e.g. in the isostructural compound Ce$_2$RhIn$_8$ \cite{218mal}). On the other hand, as no corresponding anomaly near $T_{\text{t}}$ was found for the polycrystalline sample \cite{sces08}, the observed feature may not be magnetic in origin after all. Verification of the nature of this singularity requires further studies, in particular calls for neutron diffraction experiment.

For $T_{\text{t}} < T < T_{\text{N}}$ the resistivity can be described by the formula \cite{rhoAFM}: $\rho(T)=(\rho_0+\rho_{\text{0m}})+aT+b\Delta^2\sqrt{T/\Delta}\exp(-\Delta/T)[1+(2/3)(T/\Delta)+(2/15)(T/\Delta)^2]$, in which $\rho_0$ stands for the residual resisivity due to defects, $\rho_{\text{0m}}$ is a temperature independent magnetic contribution due to spin disorder related to the CEF ground state, the linear term accounts for the behavior dominant below $T_{\text{t}}$ (see below), and the last term represents scattering the conduction electrons by magnetic excitations, characterized by a gap $\Delta$ in the spin wave spectrum. The least-squares fit of this function to the experimental data in the range from $T_{\text{N}}$ down to about 2 K (see Fig. \ref{fig:Fig2}) yielded the parameters: $\rho_0+\rho_{\text{0m}}$ = 9.7 $\mu \Omega$cm, $a$ = 1.07 $\mu \Omega$cm/K, $b$ = 0.23 $\mu \Omega$cm/K$^2$ and $\Delta$ = 4.5 K. Below $T_{\text{t}}$ the resistivity decreases with decreasing temperature down to $T_{\text{c}}$ in an almost linear manner down to $\rho_0$ interpolated to be about 4 $\mu \Omega$cm (see the upper inset to Fig. \ref{fig:Fig2}). Though the fairly narrow temperature interval $T_{\text{c}}$ -- $T_{\text{t}}$ hampers any reliable physical interpretation of the observed behavior, it is worth noting that the relation $\rho \sim T$ (with nearly the same proportionality coefficient $a$) has been established over an extended temperature range 0.35-6 K for the polycrystalline sample of Ce$_2$PdIn$_8$ and attributed to critical spin fluctuations \cite{sces08}. The absence of SC in the polycrystal studied may suggest unconventional coupling of Cooper pairs, which is known to be extremely sensitive to structural disorder, internal strains, and/or tiny changes in the composition. In this context it is enough to recall a classical example of CeCu$_2$Si$_2$ that can be superconducting or magnetic, depending on small variations in the stoichiometry \cite{geg_cecu2si2}.

\begin{figure}[h]
\includegraphics[width=0.73\columnwidth]{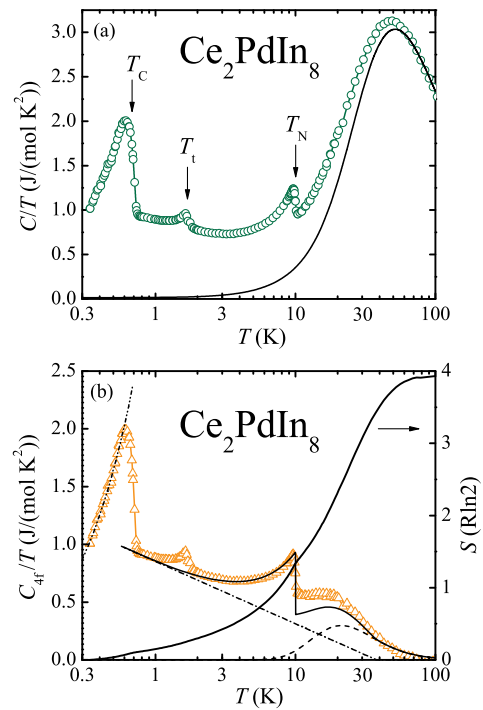}
\caption{\label{fig:Fig3} (color online) (a) Temperature dependence of the specific heat of Ce$_2$PdIn$_8$, plotted as the ratio $C/T$ on a logarithmic temperature scale. The solid line represents the phonon contribution with $\Theta_{\text{D}}$ = 184 K. (b) Temperature variations of the excess specific heat of Ce$_2$PdIn$_8$ (symbols) and the corresponding magnetic entropy (thick solid line). The dashed and dash-dotted lines are the Schottky and spin-fluctuation contributions to $C_{\text{4f}}/T$, respectively, adopted from Ref. \cite{sces08}. The thin solid curve represents the model description of the specific heat above $T_{\text{t}}$ (see text). The dash-dot-dot curve emphasizes the relation $C_{\text{4f}}/T \sim T$ in the superconducting state.}
\end{figure}

The bulk nature of all the observed phase transitions is corroborated by the thermodynamic data. A sharp peak in the specific heat of Ce$_2$PdIn$_8$, presented in Fig. \ref{fig:Fig3}a as the ratio $C/T$ vs. $T$, occurs at a temperature that perfectly matches the onset of the SC state at $T_{\text{c}}$. In turn, a clear $\lambda$-type anomaly at $T_{\text{N}}$ = 10 K corroborates the AF state. Even the faint change of slope in $\rho(T)$, observed at $T_{\text{t}}$, finds its correspondence in the form of another peak in $C/T$. In order to analyze the heat capacity in more detail, in the first step the phonon contribution was subtracted from the measured data assuming a Debye model with the characteristic temperature $\Theta_{\text{D}}$ = 184 K (see Fig. \ref{fig:Fig3}a), as derived in the study of the polycrystalline sample of Ce$_2$PdIn$_8$ \cite{sces08}. So-obtained excess specific heat $C_{\text{4f}}$, due to 4$f$ electrons, is shown in Fig. \ref{fig:Fig3}b. The magnetic entropy $S(T) = \int_{0}^{T}C_{\text{4f}}/T dT$ released by $T_{\text{N}}$ is only 0.67$R$ln2 per Ce atom , and the full value corresponding to the doublet ground state is reached near 17 K. Thus, the entropy is strongly reduced, in a manner typical for Kondo systems. The Kondo temperature estimated from $S(T)$ in terms of the Bethe Ansatz prediction for spin $s$ = 1/2 \cite{hewson} amounts to about 10 K, in perfect agreement with the value derived from the magnetic susceptibility data. Subsequently, it was assumed that the Schottky $C_{\text{Sch}}$ and spin fluctuation $C_{\text{sf}}$ contributions to the specific heat of the measured single crystal of Ce$_2$PdIn$_8$ are equal to those determined in Ref. \cite{sces08} for the polycrystal, i.e. the energy splittings between the CEF ground state and the first and second excited doublets are $\Delta_{\text{CEF1}}$ = 60 K and $\Delta_{\text{CEF2}}$ = 198 K, while the characteristic temperature of spin fluctuations in the system amounts to $T_0$ = 38 K. With these assumptions, $C_{\text{4f}}/T$ can be modeled as shown in Fig. \ref{fig:Fig3}b, where the magnon contribution below $T_{\text{N}}$ was added in the form $C_{\text{mag}} = A_{\text{C}}\Delta^4\sqrt{T/\Delta}\exp(-\Delta/T)[1+(39/20)(T/\Delta)+(51/32)(T/\Delta)^2]$, appropriate for antiferromagnets \cite{cAFM}. Here, the spin-wave gap $\Delta$ = 4.5 K was adopted, as derived from the $\rho(T)$ data, and the only adjustable parameter was $A_{\text{C}} = 12 \times 10^{-4}$ J/(mol K$^4$), a coefficient related to the spin-wave stiffness $D$ by the proportionality $A_{\text{C}} \propto 1/D^3$. As apparent from Fig. \ref{fig:Fig3}b, the model approximates the experimental data quite well, except for the region just above $T_{\text{N}}$ and below $T_{\text{t}}$. While the failure of this simple approach in the former region can be attributed to neglecting critical spin fluctuations, it may probably be improved by considering magnon contribution below the phase transition at $T_{\text{t}}$. However, because of the narrow temperature interval $T_{\text{c}}$-$T_{\text{t}}$ no attempt was made to fit the $C_{\text{4f}}/T$ data in this range.

\begin{figure}[ht]
\includegraphics[width=0.95\columnwidth]{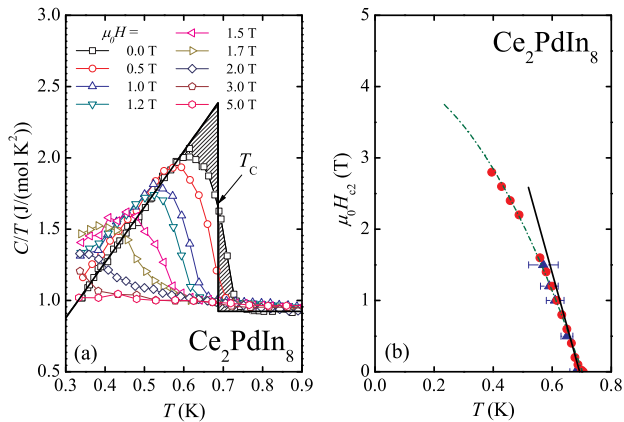}
\caption{\label{fig:Fig4} (color online) (a) Temperature dependencies of the specific heat over temperature ratio of Ce$_2$PdIn$_8$, measured in zero and a few magnetic fields applied along the $c$-axis. The hatched fields mark the equal entropy construction used for the determination of the critical temperature $T_{\text{c}}$ for the data taken in $\mu_0H$ = 0. (b) Upper critical field versus temperature determined for Ce$_2$PdIn$_8$ from the resistivity (circles) and specific heat (triangles) data. The solid line yields d$\mu_0H_{\text{c2}}$/d$T$ = -14.3 T/K. The dashed line is a guide for the eye.}
\end{figure}

In order to determine the SC transition temperature the entropy balance between the normal and SC states was checked (see Fig. \ref{fig:Fig4}a) that yielded $T_{\text{c}}$ = 0.68 K, in perfect agreement with the resistivity data. The idealized specific heat jump $\Delta C/T_{\text{c}}$ is equal to about 1.5 J/(mol K$^2$). Upon applying magnetic field $T_{\text{c}}$ decreases with raising the field strength and the SC gets suppressed in a field of 5 T. From the results presented in Fig. \ref{fig:Fig4}a, the normal state Sommerfeld coefficient $\gamma_{\text{n}}$ can be estimated as about 1 J/(mol K$^2$) per formula unit, leading to the ratio $\Delta C/\gamma_{\text{n}}T_{\text{c}} \simeq$ 1.5, i.e. close to the BCS value of 1.43. The electron-phonon coupling constant $\lambda$, calculated from the relation $\Delta C/\gamma_{\text{n}}T_{\text{c}} = 1.43+0.942\lambda^2-0.195\lambda^3$ \cite{kresin} amounts to 0.49, thus indicating that  Ce$_2$PdIn$_8$ is a weakly-coupled superconductor. As shown in Fig. \ref{fig:Fig3}b, below $T_{\text{c}}$ the ratio $C/T$ varies linearly with $T$, i.e. in a manner characteristic of unconventional superconductors with line nodes in their energy gap structure. This finding provokes an appealing question whether SC in Ce$_2$PdIn$_8$ is unconventional, like in the Ce\emph{M}In$_5$ (\emph{M} = Co, Rh, Ir) relatives \cite{heg_cerhin5,pet_cecoin5,pet_ceirin5}. To verify the latter intriguing conjecture further low-temperature studies are necessary.

The temperature dependence of the upper critical field $H_{\text{c2}}$, derived from the electrical resistivity and specific heat data, is shown in Fig.~\ref{fig:Fig4}b. The initial slope $d\mu_0H_{\text{c2}}/dT$ is as large as about -14.3 T/K, and extrapolation of the field dependent SC transition temperature towards zero gives $\mu_0H_{\text{c2}}(0) \approx$ 4.8 T. Both these enhanced values hint at heavy quasiparticles forming Cooper pairs. From the formulas given in Refs. \cite{SC_cecu2si2,orlando,hw,gl}, one may calculate a number of parameters that characterize Ce$_2$PdIn$_8$ in the SC state. All these parameters are listed in Table I, and compared with the corresponding values reported for an archetypal HF superconductor CeCu$_2$Si$_2$ \cite{SC_cecu2si2}. The obtained Ginzburg-Landau parameter $\kappa_{\text{GL}}$ = 21 clearly manifests type II superconductivity, and the relation $l \gg \xi_0$ indicates that Ce$_2$PdIn$_8$ is a clean-limit superconductor. In general, Ce$_2$PdIn$_8$ appears to be very similar to CeCu$_2$Si$_2$, except for the latter distinct feature.

\begin{table}[h]
\caption{\label{tab:table1} Comparison of the main thermodynamic parameters of Ce$_2$PdIn$_8$ and CeCu$_2$Si$_2$  \cite{SC_cecu2si2}; $k_{\text{F}}$ - Fermi wave number, $v_{\text{F}}$ - Fermi velocity, $m^*$ - effective mass, $l$ - quasiparticles mean free path, $\xi_0$ - BCS coherence length, $\xi_{\text{GL}}$ - Ginzburg-Landau coherence length, $\lambda_{\text{GL}}$ - penetration depth}
\begin{tabular}{ccc}
\\\hline
  &           Ce$_2$PdIn$_8$          &        CeCu$_2$Si$_2$ \
\\\hline
$T_{\text{c}}$ [K]                       &   0.68        &       0.64 \\
$\gamma_{\text{n}}$ [mJ/(mol$_{\text{f.u.}}$K$^2$)]       &   1000         &       1006 \\
$\mu_0 H_{\text{c2}}$ [T]                &   $\sim$ 4.8    &      $\sim$ 1.7 \\
-d$\mu_0 H_{\text{c2}}$/d$T$ [T/K]         &   14.3        &       5.8 \\
$k_{\text{F}}$ [10$^{10}$ m$^{-1}$]         &   0.92         &           1.6 \\
$v_{\text{F}}$ [10$^3$ m/s]         &   4.15         &           8.7 \\
$m^*$ [m$_0$]         &   257         &           220 \\
$\xi_0$ [nm]         &   5.4         &           19 \\
$l$ [mm]         &   42         &           12 \\
$\xi_{\text{GL}}$ [nm]         &   8.2         &           9 \\
$\lambda_{\text{GL}}$ [nm]         &   174         &           200 \\
$\kappa_{\text{GL}}$ [-]         &   21         &           22 \\
\end{tabular}
\end{table}

In summary, single-crystalline Ce$_2$PdIn$_8$ is an ambient-pressure HF superconductor, alike the closely related compounds CeCoIn$_5$  \cite{pet_cecoin5}, CeIrIn$_5$ \cite{pet_ceirin5} and Ce$_2$CoIn$_8$  \cite{che_ce2coin8}. However, in contrast to these phases, SC in Ce$_2$PdIn$_8$ emerges out of the long-range AF state. In this respect, the compound resembles antiferromagnetic CeRhIn$_5$, in which the two cooperative phenomena compete with each other at high pressures \cite{heg_cerhin5}. For the latter indide a purely SC ground state is observed for pressures higher than $p^*_c \approx$ 2 GPa, while some sort of coexistence of both orderings occurs in a limited pressure range below $p^*_c$ \cite{kne_cerhin5} or likely even down to ambient pressure \cite{maple_cerhin5}. Upon doping CeRhIn$_5$ with Co or Ir atoms, AF gets gradually suppressed and SC is observed, coexisting with the magnetic ordering in substantial ranges of dopant concentration (0.4 $\leq x \leq$ 0.6 for Co and 0.3 $< x <$ 0.6 for Ir) \cite{co_cerhin5,ir_cerhin5}. Similar behavior was established for Cd-doped CeCoIn$_5$, namely AF order coexists with SC at low temperatures for a nominal dopant concentration 0.075 $< x <$ 0.15 \cite{cd_cecoin5}. In the case of Ce$_2$PdIn$_8$, further in-depth studies are necessary to clarify whether SC in this compound has an unconventional character and if it coexists with the magnetism on a microscopic scale.

\end{document}